%% file: whitepaper-5-07.tex
\documentclass[12pt]{cernrepNum-v2}
\usepackage{calc}
\usepackage{graphicx}
\usepackage{here}
\usepackage{color}
\usepackage{amsmath}
\usepackage{longtable}
\usepackage{graphics}
\usepackage{subfigure}
\usepackage{setspace}
\newcommand{\gmt}{(g_\mu-2)}

\def\g2{{$(g-2)$} }
\def\be{\begin{equation}}
\def\ee{\end{equation}}
\def\ben{\begin{enumerate}}
\def\een{\end{enumerate}}
\def\bi{\begin{itemize}}
\def\ei{\end{itemize}}
\def\bs{\begin{slide}}
\def\es{\end{slide}}
\def\bea{\begin{eqnarray} }
\def\eea{\end{eqnarray} }
\def\bc{\begin{center} }
\def\ec{\end{center} }

\newcommand{\amu}[1][]{\ensuremath{a_{\mu^{#1}}}}
\newcommand{\gm}{\ensuremath{(g-2)}}
\newcommand{\wa}{\mbox{\ensuremath{\omega_a}}}

\begin{document}

\title{The Physics Case for the New Muon $(g-2)$ Experiment}

\author{ David W. Hertzog$^3$, James P. Miller$^1$, Eduardo de Rafael$^4$, 
B.Lee Roberts$^1$,
Dominik St\"ockinger$^{2}$
}
\institute{\ \ \ 
  \\
$^1$Department of Physics, Boston University, 
Boston, MA 02215, USA\\
$^2$Department of Physics and Astronomy, University of
Glasgow, Glasgow G12 8QQ, UK\\
$^3$Department of Physics, University of Illinois, 
Urbana-Champaign, IL 61801, USA\\
$^4$Centre de Physique Th\'eorique, CNRS-Luminy, Case 907, 
F-13288 Marseille Cedex 9, France}
\vskip0.5in

\maketitle
                     
\singlespacing                                                           

\begin{abstract}

This White Paper briefly reviews the present status of the muon \g2
experiment and the physics motivation for a new effort.  The present
comparison between experiment and theory indicates a tantalizing
$3.4~\sigma$ deviation.  An improvement in precision on this
comparison by a factor of 2---with the central value remaining
unchanged---will exceed the ``discovery'' threshold, with a
sensitivity above $6~\sigma$.  The 2.5-fold reduction improvement
goal of the new Brookhaven E969 experiment, along with continued
steady reduction of the standard model theory uncertainty, will
achieve this more definitive test.

Already, the \g2 result is arguably the most compelling indicator of
physics beyond the standard model and, at the very least, it
represents a major constraint for speculative new theories such as
supersymmetry or extra dimensions. In this report, we summarize the
present experimental status and provide an up-to-date accounting of
the standard model theory, including the expectations for improvement
in the hadronic contributions, which dominate the overall
uncertainty.  Our primary focus is on the physics case that motivates
improved experimental and theoretical efforts.  Accordingly, we give
examples of specific new-physics implications in the context of
direct searches at the LHC as well as general arguments about the
role of an improved \g2 measurement. A brief summary of the plans for
an upgraded effort complete the report.  
\\
\\

\end{abstract}
\newpage
%
\input{Introduction.tex}

%
\input{StandardModel.tex}
\input{BeyondSM.tex}

\input{improvements.tex}

\input{Summary.tex}
%
%

\input{Bibliography.tex}
%
%
%
%
\end{document}

%% file: Introduction.tex
\section{Introduction}

\begin{figure}[h!]
\begin{center}
\begin{minipage}{\textwidth-.2in}
  \begin{minipage}{\textwidth/2-0.4in}
\includegraphics[width=0.9\textwidth,angle=0]{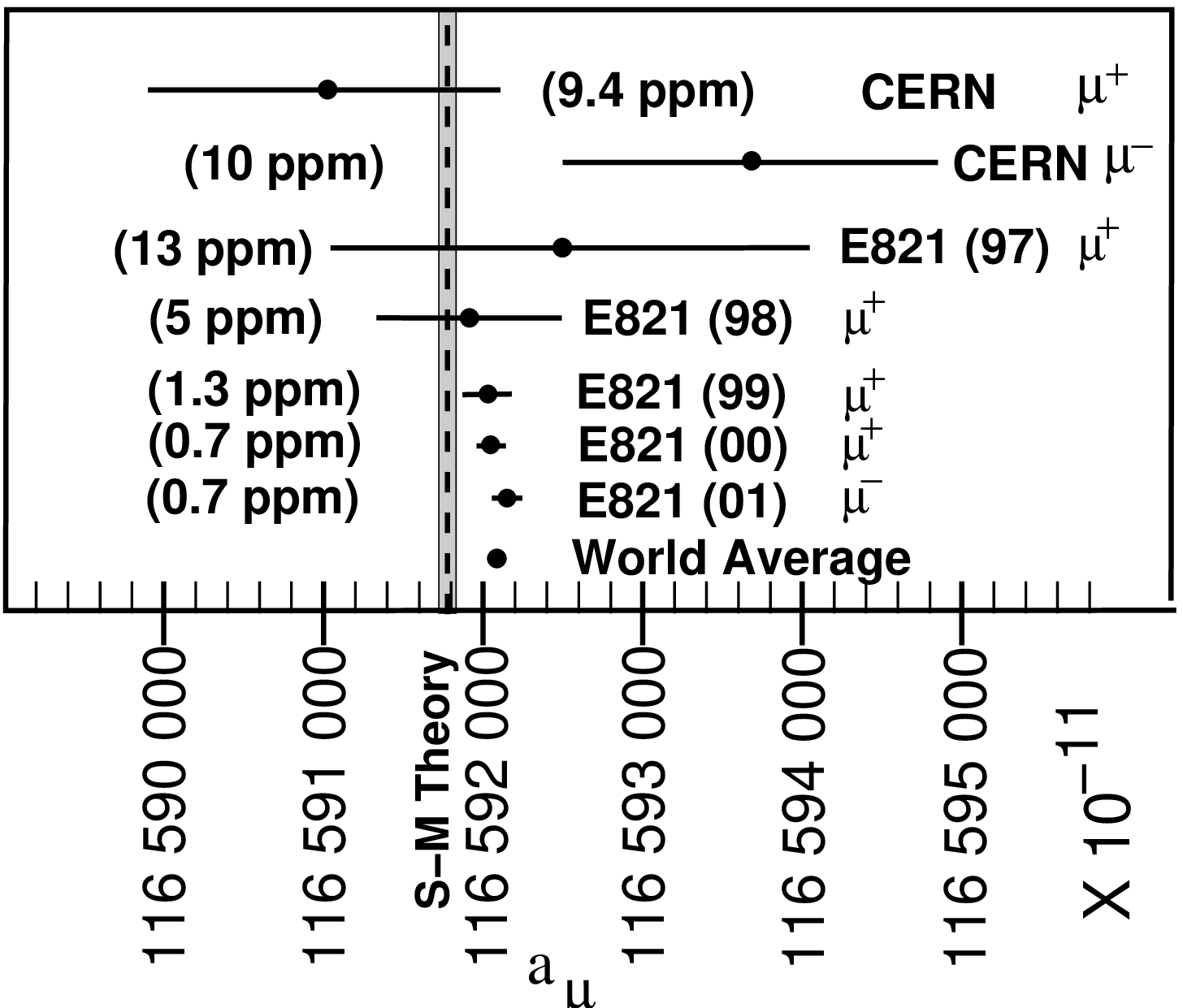}
  \end{minipage}
\hskip0.2in
  \begin{minipage}{\textwidth/2}
    \begin{center}
      \begin{tabular}{lclr} \hline
          Year& Polarity & $a_\mu \times 10^{10}$ & $\sigma \amu$ [ppm] \\
          \hline
          1997 & ~$\mu^+$~ &   11\,659\,251(150) & 13  \\
          1998 & $\mu^+$ &   11\,659\,191(59) & 5 \\
          1999 & $\mu^+$ &   11\,659\,202(15) & 1.3  \\
          2000 & $\mu^+$ & 11\,659\,204(9) & 0.7  \\
          2001 & $\mu^-$ & 11\,659\,214(9) & 0.7  \\
          \hline
          Avg. & & 11\,659\,208.0(6.3) & 0.54  \\
        \hline 
        \\
        \\
      \end{tabular}
    \end{center}
  \end{minipage}
\end{minipage}
\end{center}
\caption{Measurements of the muon anomaly, indicating the value, as well as
the muon's sign. To obtain
the value of $a_{\mu^-}$, as well as the world average,
 {\sl CPT} invariance is assumed. The theory value is taken from
Ref.~\cite{MdRR2007}, which uses electron-positron
annihilation to determine the hadronic contribution.
\label{fg-tb:g2results}}
\end{figure}

The anomalous magnetic moment of the muon, $a_\mu = (g-2)_\mu /2$ 
has been measured in 
 Experiment E821 at the Brookhaven AGS, which
 completed data collection in 2001.  All results are 
published~\cite{carey,brown1,brown2,bennett1,bennett2}, and the
measurements are  shown in Fig.~\ref{fg-tb:g2results}.
A comprehensive summary of the experiment, containing many of the
details of the methods used for data collection and analysis, was
published~\cite{bennett3} in 2006, and general reviews of the
experiment and theory are also
available~\cite{HertzogMorse2004,FS,davmar,MdRR2007}.

The nearly equally precise experimental determinations of $a_\mu$
 from positive and negative muon
data samples can be combined under the assumption of {\sl CPT}
invariance to give
\be
\amu ^{( \rm Exp)} = 116\,592\,080(63) \times 10^{-11}.
\label{eq:exp}
\ee
The final error of
0.54~ppm consists of a 0.46~ppm statistical component
and a 0.28~systematic
component, combined in quadrature. 
An upgraded version of E821, E969~\cite{E969}, has been proposed and 
received scientific 
approval at Brookhaven, with the goal of reducing
 the experimental uncertainty on $a_\mu$ by a factor
 of 2.5, down to $ \pm 25 \times 10^{-11}$.

In 2007 the standard model prediction for $a_\mu$ was updated based on 
new results from the CMD-2
 and SND $e^+e^-$ annihilation experiments,
and from radiative return measurements from BaBar. Additional theory
work on hadronic light-by-light scattering was also completed.  In
the review of Ref.~\cite{MdRR2007}, the most recent theory value is
determined to be 
\be
 a_{\mu}^{{\rm ( SM)}} = 116\,591\,785 (61) \times 10^{-11}.
\label{eq:theory}
\ee
The uncertainty of 0.52~ppm is close to the error on the
experimental value,
with a difference between the standard model and experiment of
\be
\Delta a_\mu ^{(\rm today)} = a_{\mu}^{{\rm (Exp)}} -  a_{\mu}^{{\rm ( SM)}}
 = (295 \pm 88)\times10^{-11},
\label{eq:Delta}
\ee
a 3.4 standard deviation difference.

%

%% file: StandardModel.tex
\section{The Standard-Model Value of the Anomaly}

\begin{figure}[h!]
\begin{center}
  \includegraphics[width=0.8\textwidth,angle=0]{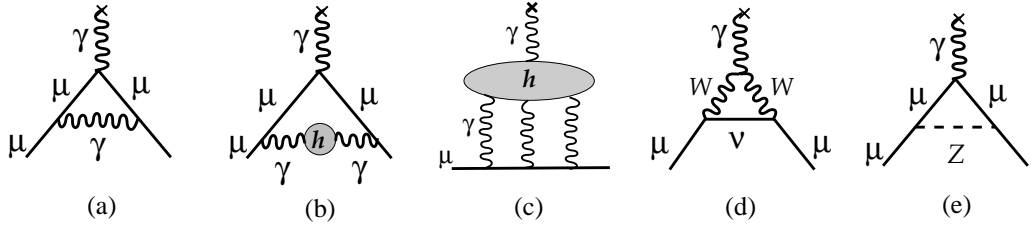}
\end{center}
  \caption{The Feynman graphs for: (a) Lowest-order QED (Schwinger term);
 (b) Lowest-order hadronic
contribution; (c) The hadronic light-by-light contribution; 
(d)-(e) the lowest order electroweak $W$ and $Z$ contributions.
  With the present 
limits on $m_H$, the contribution from the single Higgs loop is negligible.}
  \label{fg:radcor}
\end{figure}

The standard model value of a lepton's anomaly,
$a_\ell$, has  contributions from three different sets of radiative
processes: 
\bi
\item quantum electrodynamics (QED) -- with loops containing 
leptons ($\ell = e,\mu,\tau$) and photons; 
\item hadronic -- with  hadrons in vacuum polarization
 loops;
\item weak -- with loops involving the  bosons $W,Z,$ and Higgs.
\ei
Examples  are shown in Fig.~\ref{fg:radcor}.  Thus
\begin{equation}
a_{\mu}^{{\rm ( SM)}}
= a_{\mu}^{({\rm QED})} + a_{\mu}^{({\rm hadronic})} +
a_{\mu}^{({\rm EW})}\, ,
\end{equation}
with the uncertainty dominated by the hadronic term.
The standard-model value of the muon anomaly has recently been 
reviewed~\cite{MdRR2007}, and the latest values of the contributions 
are given in Table~\ref{tb:SM06}.  

\begin{table}[h]
\begin{center} \caption{Standard-model contributions to the muon
anomalous magnetic dipole moment, $a_\mu$. All values are taken 
from Ref.~\cite{MdRR2007}.Possible improvements in errors 
on the hadronic corrections are also listed. As additional 
$e^+e^-$ data become available, the error listed for radiative corrections
will decrease. }
\label{tb:SM06}
{\small 
\begin{tabular}{|l|c|l |}
\hline
{\em {\textbf {Effect}}} & {\em {\textbf { Contribution }}$\times 10^{11}$} 
& {\em {\textbf { Future}}} $\mathbf \sigma$\\
\hline
\hline
QED  & $({116~584~718.09\pm 0.14_{5\rm loops}\pm 0.08_{\alpha}
\pm 0.04_{\rm masses})\times 10^{-11}} $ & \\
\hline
Hadronic (lowest order)  &$a_{\mu}^{(\rm HVP;1)}=(6901 
\pm 42_{\mbox{\rm  exp}}
\pm 19_{\mbox{\rm  rad}} \pm 7_{\mbox{\rm  QCD}})$ 
& $\pm 30_{\mbox{\rm  exp}}
 \pm  8_{\mbox{\rm rad}}  \pm 7_{\mbox{\rm  QCD}}$   \\
Hadronic (higher order)  & 
$a_{\mu}^{(\rm HVP;h.o.)}=\left(-97.9\pm 0.9_{\mbox{\rm exp}} 
\pm 0.3_{\mbox{\rm  rad}}\right)$ & \\
Hadronic (light-by-light) & $a_{\mu}^{(\rm HLLS)}=(110\pm 40)$
& $  16.5$  \\
\hline
Electroweak & $a_{\mu}^{(\rm EW)}=(154\pm 2_{\rm  M_H}  \pm1_{\rm had} )$ & \\
\hline
\end{tabular}
}
\end{center}
\end{table}

 The dominant contribution from quantum electrodynamics (QED),
called the Schwinger term~\cite{Sch48}, is $a^{({\rm QED;2})} = \alpha/2 \pi$,
which is shown diagrammatically in Fig.~\ref{fg:radcor}(a).
The QED contributions have been calculated through four loops, (eighth-order 
or $[\alpha/\pi]^4$) with the
leading five-loop (tenth-order) contributions estimated~\cite{kno06b}. 
It is the uncertainty on the tenth-order contribution that
dominates the error on $a_\mu^{(\rm QED)}$ given in Table~\ref{tb:SM06}. 

The one-loop electroweak contributions were calculated
shortly after the electroweak theory  was shown to be renormalizeable.
It has now been calculated through two loops.  The leading-log three-loop
effects have been estimated and found to be negligible.  The electroweak
contribution through two loops is 
given in  Table~\ref{tb:SM06}.

The hadronic contribution cannot be calculated
 from perturbative QCD only because of the low-energy scales involved,
and the uncertainty on $a_{\mu}^{(\rm HVP;1)}$   dominates
the total uncertainty on the standard-model value.
There are three distinct components in 
the hadronic contribution:
\bi
\item The lowest-order hadronic vacuum polarization, as shown 
in Fig.~\ref{fg:radcor}(b).

\item Higher-order hadronic vacuum polarization loops 
 (excluding the hadronic light-by-light term).

\item The hadronic light-by-light term, shown in Fig.~\ref{fg:radcor}(c).

\ei

Dispersion theory relates the (bare) cross section for 
$e^+ e^- \rightarrow {\rm hadrons}$
to the lowest-order hadronic contribution to $a_\mu$,
\begin{equation}
a_{\mu}^{({\rm hadronic;1})}=\left(\frac{\alpha m_{\mu}}{ 3\pi}\right)^2
\int^{\infty} _{4m_{\pi}^2} \frac{ds }{ s^2}K(s)R(s)~,
\quad {\rm where} \quad
R\equiv \frac{ {\sigma_{\rm tot}(e^+e^-\to{\rm hadrons})}}{
\sigma_{\rm tot}(e^+e^-\to\mu^+\mu^-)}
\label{eq:dispersion}
\end{equation}
and $K(s)$ is a known function~\cite{MdRR2007}. 
Experimental data are used as input
for $R(s)$~\cite{Davier06,Hagiwara_06}.
 The only assumptions here are analyticity and the optical
theorem.
 The factor of $s^{-2}$ in
the dispersion relation means that the $\rho$ resonance region dominates
the dispersion integral, with the region up to $\sqrt s \sim 2$~GeV being
most important.  As indicated in Table~\ref{tb:SM06}, the uncertainty on
$a_{\mu}^{(\rm HVP;1)}$ consists of three parts; the first two are associated
with the numerical integration of the $(e^+ e^- \rightarrow {\rm hadrons})$
data.  Some of these data are quite old, and 
the uncertainty because of missing radiative corrections is indicated by 
$\sigma_{\rm rad}$~\cite{davierpc,Davier06}.  
As more $e^+e^-$ data become available, both of
these errors will improve.  A reasonable expectation 
is that $\sigma_{ exp}$ will improve by at least $\sqrt 2$.  
As new data which  include all radiative corrections replace the old data,
$\sigma_{\rm rad}$ will become much less important.

It has been proposed that the hadronic contributions 
could also be determined from hadronic $\tau$-decay data, using the conserved
vector current (CVC) hypothesis.  Such an approach can only give the
isovector part of the amplitude, e.g. the $\rho$ but not the $\omega$ 
intermediate states. In contrast, the  $e^+e^-$ annihilation
cross section contains both isovector and isoscalar contributions,
with the cusp from $\rho-\omega$ interference as a dominant feature.
Since hadronic $\tau$ decay goes through
the charged $\rho$ resonance, and $e^+e^-$ annihilation goes through
the neutral $\rho$, understanding
the isospin corrections is essential to this approach.  This use of 
CVC can be checked by comparing the hadronic contribution to $a_\mu$ obtained
from each method.  Alternately, one can take
 the measured branching ratio for
 $\tau^- \rightarrow V^- \nu_\tau $, where $V$ is any vector final
state (e.g. $\pi^- \pi^0$) and compare it to that predicted using
CVC and $e^+e^-$ data,  applying all the appropriate 
isospin corrections. At present, neither comparison
gives a satisfactory result.  For example, 
the differences between the measured
branching ratios, and those obtained from CVC are
4.5~$\sigma$ and 3.6~$\sigma$ for the $\pi^- \pi^0$ and the $3\pi \pi^0$
channels respectively, with good agreement being
obtained for the $\pi 3\pi^0$ channel~\cite{Davier06}.
At present the prescription of CVC with the appropriate isospin correction
 seems to have aspects that are not understood.
 Given two consistent
$e^+e^-$ data sets and the uncertainties inherent in the
 required isospin corrections to the $\tau$ data,
the most recent standard-model
 evaluations do not use the $\tau$ data to
determine $a^{(\rm hadronic;1)}$~\cite{Davier06,Hagiwara_06,MdRR2007}.
We return to this point in the next section.

The sum of the QED, hadronic and electroweak
contributions in Table~\ref{tb:SM06}, 
adding the
errors in quadrature, gives the standard model value in Eq.~\ref{eq:theory}.
When  compared with the experimental world 
average~\cite{bennett2,bennett3} in Eq.~\ref{eq:exp},
one finds the $3.4~\sigma$ difference 
given in Eq.~\ref{eq:Delta}.  This difference 
 is at the ``interesting'' level, and makes it clear
that further work should be done to clarify whether there is a true
discrepancy.   It is estimated that the theory
could improve by a factor of two~\cite{MdRR2007,Eid}, as could the experiment.

\subsection{Expected Improvements in the Standard-Model Value}

Over the past fifteen years, significant progress has been made in improving
the standard model value of $a_\mu$.  The QED and weak contributions
are now very well known, and substantial progress has been made 
on the hadronic contribution because of the large quantity of 
new high-quality
 $e^+e^-$ data, both on the $2 \pi$ final state from
Novosibirsk~\cite{Akhmetshin_06a,Akhmetshin_06b,Achasov_06}, and
on multi-hadron final states 
from BaBar~\cite{Aubert04,Aubert05,Aubert06,Aubert07}. 
The BaBar detector
operates at a fixed beam energy, and final-state
 hadrons are in coincidence with an initial-state photon 
which lowers the center of mass energy of the collision
(often called radiative return or initial state radiation).  The KLOE
experiment at Frascati has published data on the $\pi \pi$ channel 
using radiative return~\cite{KLOE}, with the initial state radiation not
detected. The KLOE collaboration is now analyzing their data at large angles
where the soft photon is detected, and are also determining the
$\pi \pi / \mu \mu$ ratio where many systematic effects cancel.
In the next year, the BaBar collaboration will
release their $2\pi$ data~\cite{davierpc} using radiative return.  
If these BaBar data confirm 
the Novosibirsk data, it will significantly increase our confidence
in our knowledge of the lowest-order contribution to 
$ a_{\mu}^{({\rm   hadronic})}$. 
In the longer term, an upgraded collider, VEPP2000 will come on line
at Novosibirsk, with the upgraded detectors CMD-3 and SND.  This
new facility will permit 
improved measurement of the annihilation cross section from
threshold up to 2.0~GeV, and will complement the data from BaBar that are
expected to be available in the next year.

The other hadronic issue is the hadronic light-by-light contribution, shown in 
Fig.~\ref{fg:radcor}(c), which has a 36\% relative 
uncertainty\cite{MdRR2007,BP07}.
So far, the only rigorous QCD result in this domain comes from the 
observation~\cite{KNPdeR02} that in the QCD large number of colors
 ($N_c$) limit, and to leading-order in the chiral expansion, the
 dominant contribution can be calculated analytically and gives
 a positive contribution. Unfortunately, to go beyond that, one 
has to resort to hadronic models. The dynamics of the models, 
however, is very much constrained by the behavior that QCD 
imposes in various limits: the chiral limit, but also the operator
 product expansion for certain kinematic configurations. 
 In other kinematic regimes, the models can also be related to observed 
processes.
 A combined effort from theorists along the lines started 
in Refs.~\cite{KNPdeR02,KNb01,MV05,BP07} could significantly reduce 
the present uncertainty. Following this line, a goal of 15\%
accuracy in hadronic light-by light determination seems possible.

We consider several potential improvements to the hadronic contribution,
and calculate the projected future significance assuming the
reduced uncertainty on  $\Delta a_\mu$,
along with the value $\Delta a_\mu^{(\rm today)}$ 
of $ 295 \times 10^{-11}$.  Four 
different scenarios are considered, with the improvements obtained
given in Table~\ref{tab:SM-imp}.

\newpage
\ben 
\item {\em Lowest-order Hadronic (L-O-Hadronic)}: 
The error on the lowest-order 
hadronic contribution contains three
pieces (see Table~\ref{tb:SM06}). We take $\sigma_{\rm exp}$
to be improved by $\sqrt 2$, and  $\sigma_{\rm rad}$ to be improved
from $19$ to $8 \times 10^{-11}$.

\item {\em Hadronic Light-by-Light (H-L-b-L)}: The  hadronic light-by-light
contribution is improved to a 15\% relative
error, and the lowest-order hadronic has the present value. 
\item {\em L-O-Hadronic and H-L-b-L}: 
Both the lowest-order and the light-by-light errors
are improved by the amount mentioned above. 
\item {\em Most Optimistic}: For the various new-physics
 calculations given below, a 
combined theory plus experiment error of  $39 \times 10^{-11}$ was
assumed 
\een

\begin{table*}[h] \caption{Potential Improvements to $a_\mu^{\rm SM}$,
and the statistical significance if
$\Delta a_\mu  = 295  \times10^{-11}$.  For each of the future scenarios,
the experimental error is assumed to be $\pm 25 \times 10^{-11}$.
\label{tab:SM-imp}}
\begin{tabular}{|l|c|c|c| } \hline
Assumption& SM Error  & Combined Error,  & 
\ $ {\underline {295 \times 10^{-11}}} $ \   \\
          &  $(\sigma_{\rm SM}\times 10^{11})$ 
&  $(\sigma_{\rm tot} \times 10^{11})$ &${\sigma_{\rm tot}}$\\
\hline
{\em Present Errors} & 61 & 88  & 3.4 \\
\hline
{\em L-O-Hadronic  } &  54 &   59   & $>5$ \\
{\em H-L-b-L} & 50&    55&  $>5$ \\
{\em Both L-O-Hadronic and H-L-b-L}\ \  & 40 &47 &$> 6$ \\
{\em Most Optimistic} & 30 & 39 &$ >7$ \\
\hline
\end{tabular}
\end{table*}

%% file: BeyondSM.tex
\section{$\gmt$ and New Physics in the Era of LHC}

The next decade will constitute a very promising and exciting era in
particle physics. Experiments at the LHC will explore physics at the
TeV-scale, an energy scale that has not been probed directly by any
previous experiment and which appears to be a crucial energy scale in
particle physics. It is linked to electroweak symmetry breaking, and
e.g.\ naturalness arguments indicate that radically new theoretical concepts
such as supersymmetry or extra space dimensions might be realized at
the TeV-scale. Furthermore, the properties of cold dark matter are
compatible with weakly interacting particles with weak-scale/TeV-scale
masses, and Grand Unification prefers the existence of supersymmetry
at the TeV-scale. Hence, there is a plethora of possibilities, and it
is likely that a rich spectrum of discoveries will be made at the
TeV-scale. In any case, experimental data at the TeV-scale should
give answers to many fundamental questions and lead to tremendous
progress in our understanding of particle physics.

Clearly, due to the expected importance and complexity of physics at the
TeV-scale, we need to combine and cross-check information from the LHC
with information from as many complementary experiments as possible.
The measurement of the muon magnetic moment $\gmt/2=\amu$ is indispensable in
this respect.

\begin{figure}[h]
\begin{center}

\includegraphics[width=0.8\textwidth]{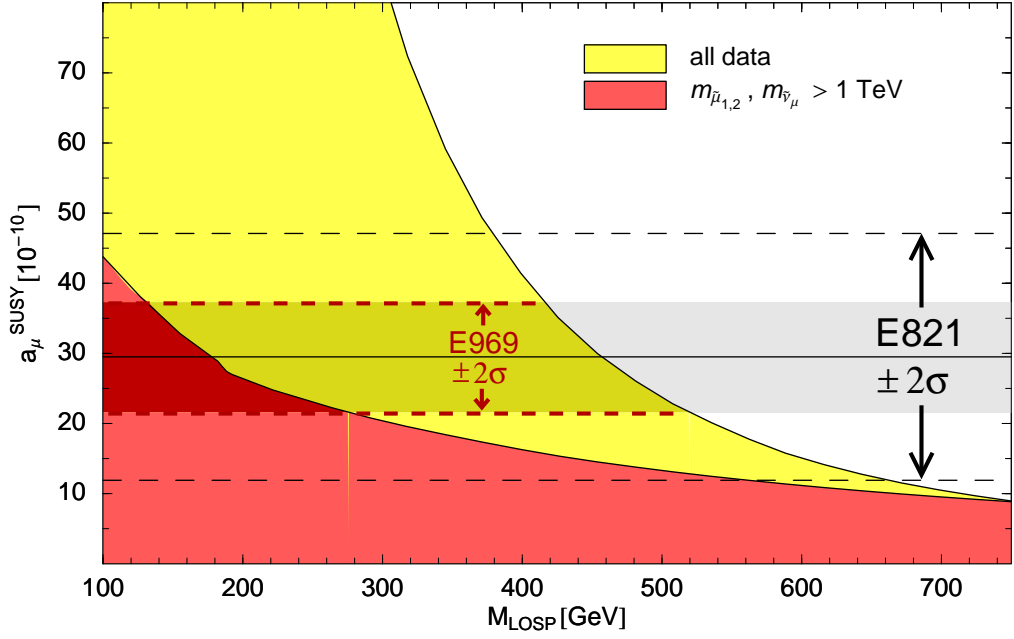}

\caption{Possible values of $\Delta\amu^{\rm MSSM}\equiv\amu^{\rm MSSM}-\amu^{\rm
    SM}$ as a function of the lightest observable supersymmetric particle mass
$M_{\rm LOSP}$. The future constraint assumes
$\Delta\amu^{(\rm future)}=295(39)\times 10^{-11}$.
The black horizontal dashed line
is the  present (E821) $2\sigma$ band; the dark-red lines give
 the future (E969) $2\sigma$ band.
The yellow region corresponds to all data
points compatible with constraints from $M_h$, $\Delta \rho$ and $b$-decays.
The red region is for smuons and sneutrinos heavier than 1 TeV.
The figure shows that the future $\amu$
measurement significantly constrains the MSSM parameter space and leads to
upper and lower mass bounds on supersymmetric particles. This would even be
the case if the future central value $\Delta\amu^{(\rm future)}$ would be
smaller than today's. For more details on the plot see Ref.~\cite{DSreview}.
\label{fig:SUSYscan} }
\end{center}
\end{figure}

The muon magnetic moment is one of the most precisely measured
and calculated quantities in elementary particle physics. Moreover,
the current experimental value of $\amu$ shows one of the largest
deviations of any observable from the corresponding standard-model
prediction (see Eq.~\ref{eq:Delta}), 
$\Delta\amu^{\rm today}=295(88)\times10^{-11}$. Owing to 
this precision $\amu$ is not only a sensitive test of all standard
model interactions, but also of possible new physics at and above the
electroweak scale. If the precision of $\Delta\amu$ is improved to
$39\times10^{-11}$, $\amu$ will be a highly sensitive probe of physics
beyond the standard model up to the TeV-scale. Results from a model-independent
MSSM parameter scan shown in
Fig.~\ref{fig:SUSYscan}
exemplifies this sensitivity for the case of the minimal supersymmetric
standard model (MSSM) by comparing the current value $\Delta\amu^{(\rm today)}$
and the future precision with the values for $\Delta\amu$ compatible with the
MSSM. 

In the following, the importance of the $\amu$ measurement in the era of
TeV-scale  physics, and particularly its usefulness as a complement to LHC, is
discussed. The discussion is centered around the following aspects:
\begin{itemize}
\item The measured value $\Delta\amu$ constitutes a definite
  benchmark that any model of new physics has to satisfy.
\item $\amu$ is particularly sensitive to quantities that are
  difficult to measure at the LHC.
\item $\amu$ is an inclusive measure of quantum effects from all
  particles.
\item $\amu$ is a very clean observable.
\item $\amu$ is a simple and beautiful quantity.
\end{itemize}

\subsection{$\amu$ as a benchmark for models of new physics}

It has been established that the LHC is sensitive to virtually all proposed
weak-scale extensions of the standard model, ranging from supersymmetry to
extra dimensions, little Higgs models and others. However, even if the
existence of physics beyond the standard model is established, it will be far
from easy for the LHC alone to identify which of the possible alternatives is
realized. The measurement of $\amu$ to $39\times 10^{-11}$ will be highly
valuable in this respect since it will provide a benchmark and stringent
selection criterion that can be imposed on any model that is tested at the
LHC. For example, if $\Delta\amu$ persists to be as large as it is today, many
non-supersymmetric models will be ruled out. If $\Delta\amu$ turns out to be
rather small, supersymmetric models will be seriously constrained. 

One example, where the power of $\amu$ as a selection criterion becomes
particularly apparent, is the distinction between the minimal supersymmetric
standard model (MSSM) and a
Universal Extra Dimension (UED) model. Both models predict the existence of
heavy partners (superpartners or Kaluza-Klein modes) of the standard-model
particles. The quantum numbers and --- for suitable model parameters --- also
the mass spectra of these heavy partners are
the same in both models. Ref.~\cite{SmillieWebber} 
analyzed whether the two models can be distinguished at
the LHC by considering spin-sensitive observables. The answer turned out to be
affirmative, however even in the collider-friendly case that the mass-spectrum
of the MSSM reference point SPS1a is realized, the separation of the two models
is not an easy task. A precise measurement of $\amu$ would be of great help in
this respect. The values predicted by the MSSM\cite{DSreview}  
 and by the UED\cite{AppelqDob} model (in the
parameter point of Ref.\ \cite{SmillieWebber}) are very 
different:\footnote{The UED result is obtained from Eq. (3.8) of reference
          \cite{AppelqDob} for the case of one extra dimension,
          $\delta=1$.}
\be
\Delta\amu^{\rm MSSM} = 298\times 10^{-11}~; 
\quad
\Delta\amu^{\rm UED} \approx
 -13\times 10^{-11}~.  
\ee
Hence, the future $\amu$ measurement would separate the two models by more
than 7 standard deviations and thus allow for a clear decision in favor of
one of the two models.

A second example is the distinction between two different, well-motivated
scenarios of supersymmetry breaking such as anomaly-mediated and
gravity-mediated supersymmetry breaking. The two scenarios lead to different
values of superpartner masses, but a major qualitative difference is the
different sign for the MSSM-parameter $\mu$ preferred in both scenarios (if
the current experimental constraint on BR$(b\to s\gamma)$ is imposed). The LHC
is not particularly sensitive to sign$(\mu)$, and thus cannot test this
fundamental difference. However, sign$(\mu)$ determines the sign of the
supersymmetry contributions to $\amu$. The current value
$\Delta\amu^{(\rm today)}=295(88)\times 10^{-11}$ already
favors sign$(\mu)=+$, but the magnitude of the
uncertainty does not allow a definite conclusion. An improved measurement with
uncertainty $39\times 10^{-11}$ has the potential to unambiguously determine
sign$(\mu)$ and thus one of the central supersymmetry parameters. Depending on
the future central value of $\Delta\amu$, either anomaly- or gravity-mediated
supersymmetry breaking could be ruled out or at least seriously constrained by
$\amu$.

\begin{figure}[h!]
\begin{center}
\subfigure[ ] {\includegraphics[width=.33\textwidth]{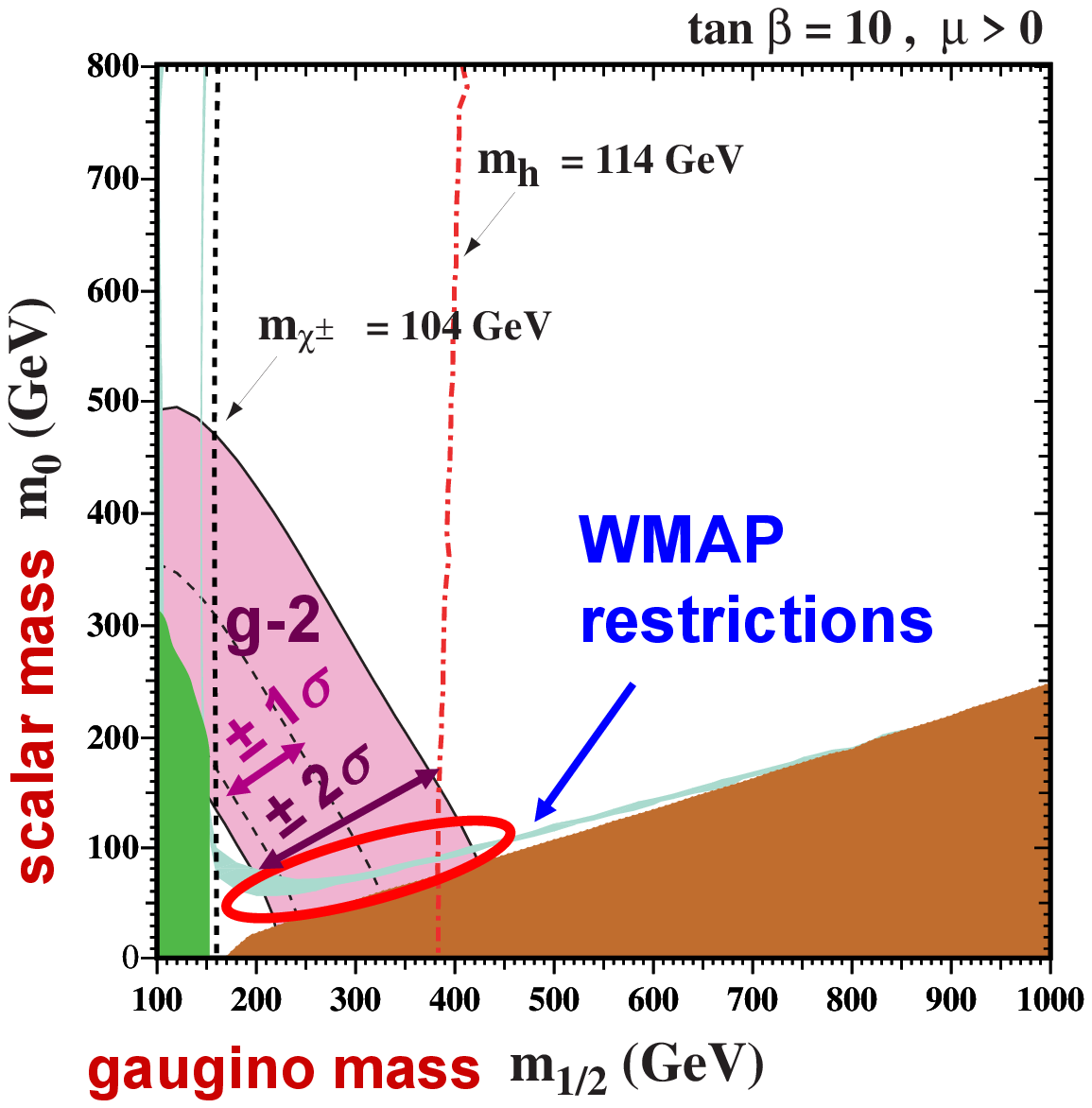}}
\subfigure[ ] {\includegraphics[width=.33\textwidth]{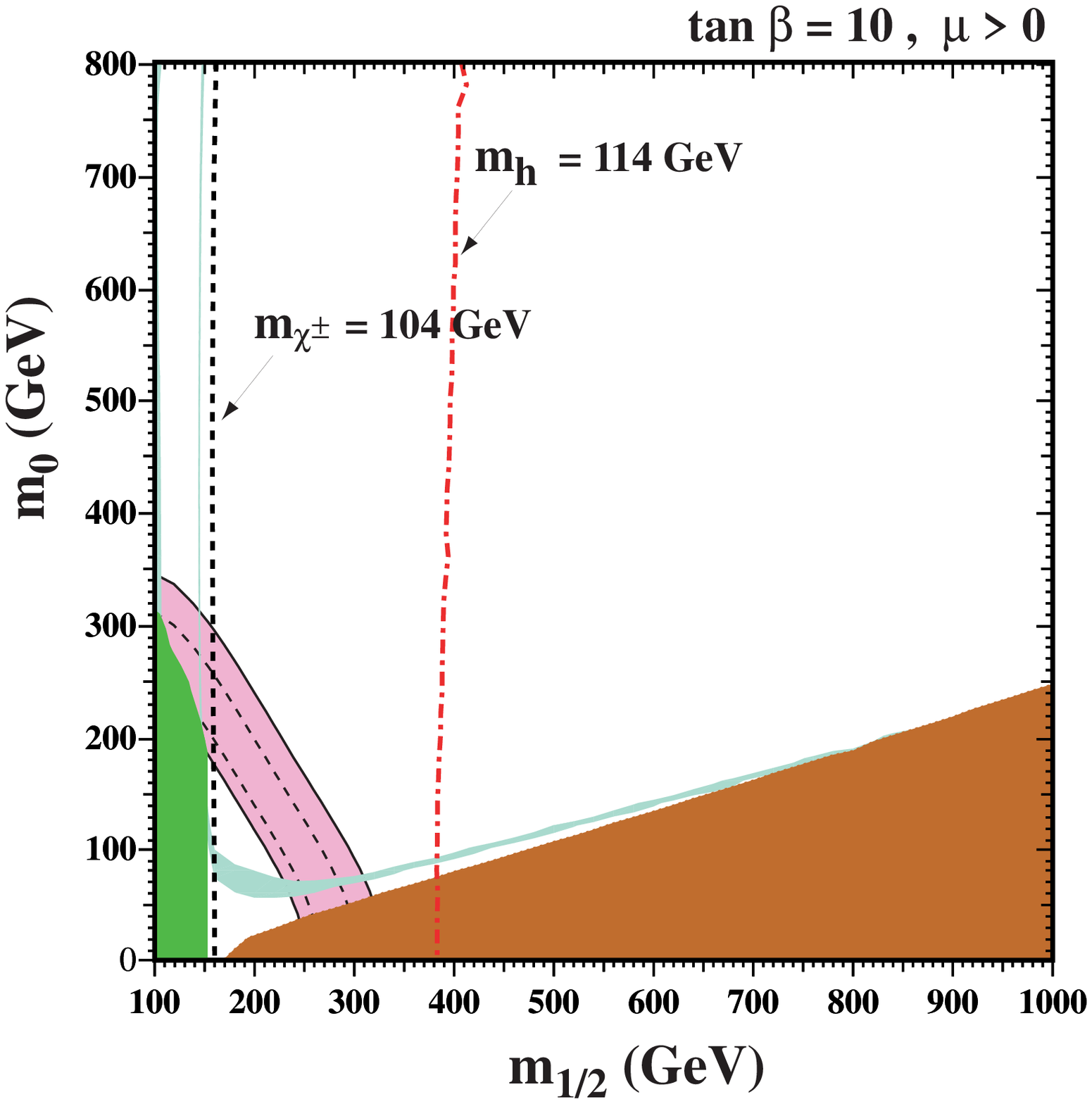}}\subfigure[]{\includegraphics[width=.33\textwidth]{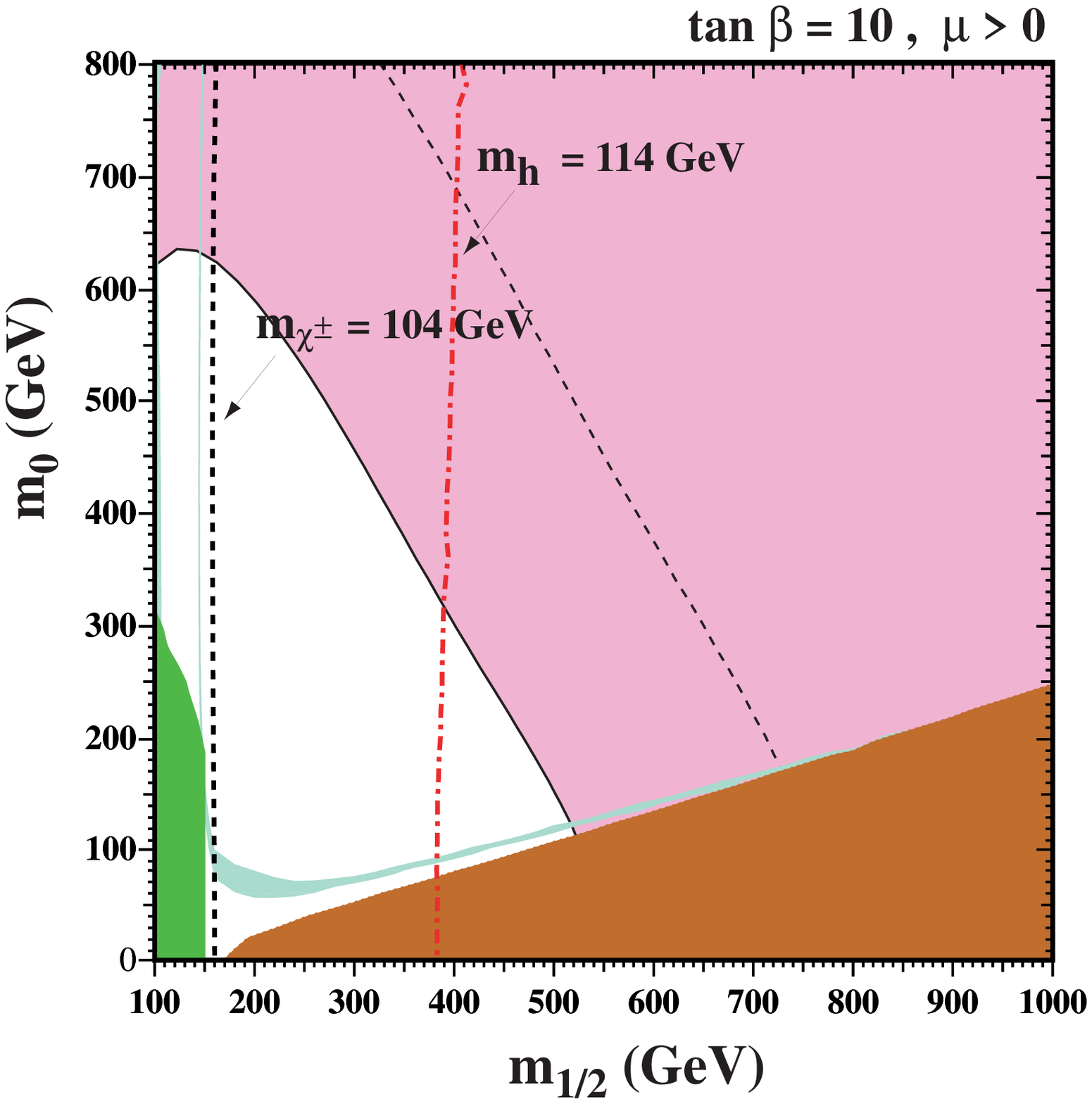}} \caption{
The $m_0$--$m_{1/2}$ plane of the CMSSM parameter space for
  $\tan\beta=10$, $A_0=0$, sign$(\mu)=+$. 
(a) The $\Delta a_{\mu}^{(\rm today)}=295(88)\times 10^{-11}$
 between experiment and standard-model theory is from
Ref.~\cite{MdRR2007}, see text. 
The brown wedge on the lower right is excluded by
the requirement the dark matter be neutral.  Direct limits on
 the Higgs and chargino $\chi^\pm$ masses are indicated by vertical
lines. Restrictions from the WMAP satellite data are shown as a 
light-blue line. The $(g-2)$ 1 and 2-standard deviation boundaries are shown
in purple.
The region ``allowed'' by WMAP and $(g-2)$  is indicated by the ellipse,
which is further restricted by the limit on $M_h$.
 (b) The plot with
 $\Delta a_{\mu}=295(39)\times 10^{-11}$, which assumes that both
the theory and experimental errors decrease to $22 \times 10^{-11}$.
(c) The same errors as (b), but $\Delta = 0$.
 (Figures courtesy of K. Olive)
\label{fg:dark}
}  
\end{center}
\end{figure}

\begin{figure}[h!]
\begin{center}
\subfigure[ ] {\includegraphics[width=.33\textwidth]{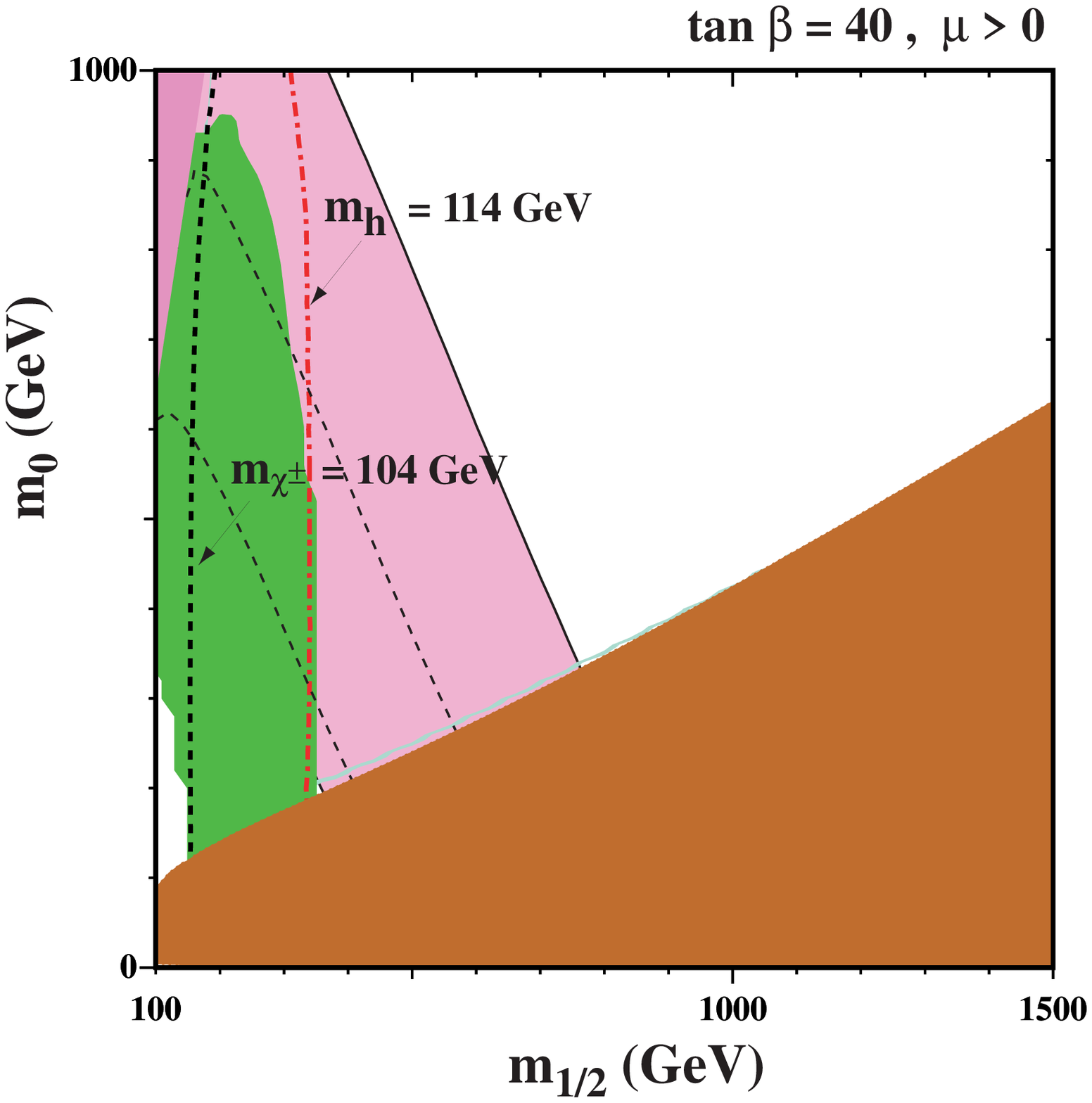}}
\subfigure[ ]
{\includegraphics[width=.33\textwidth]{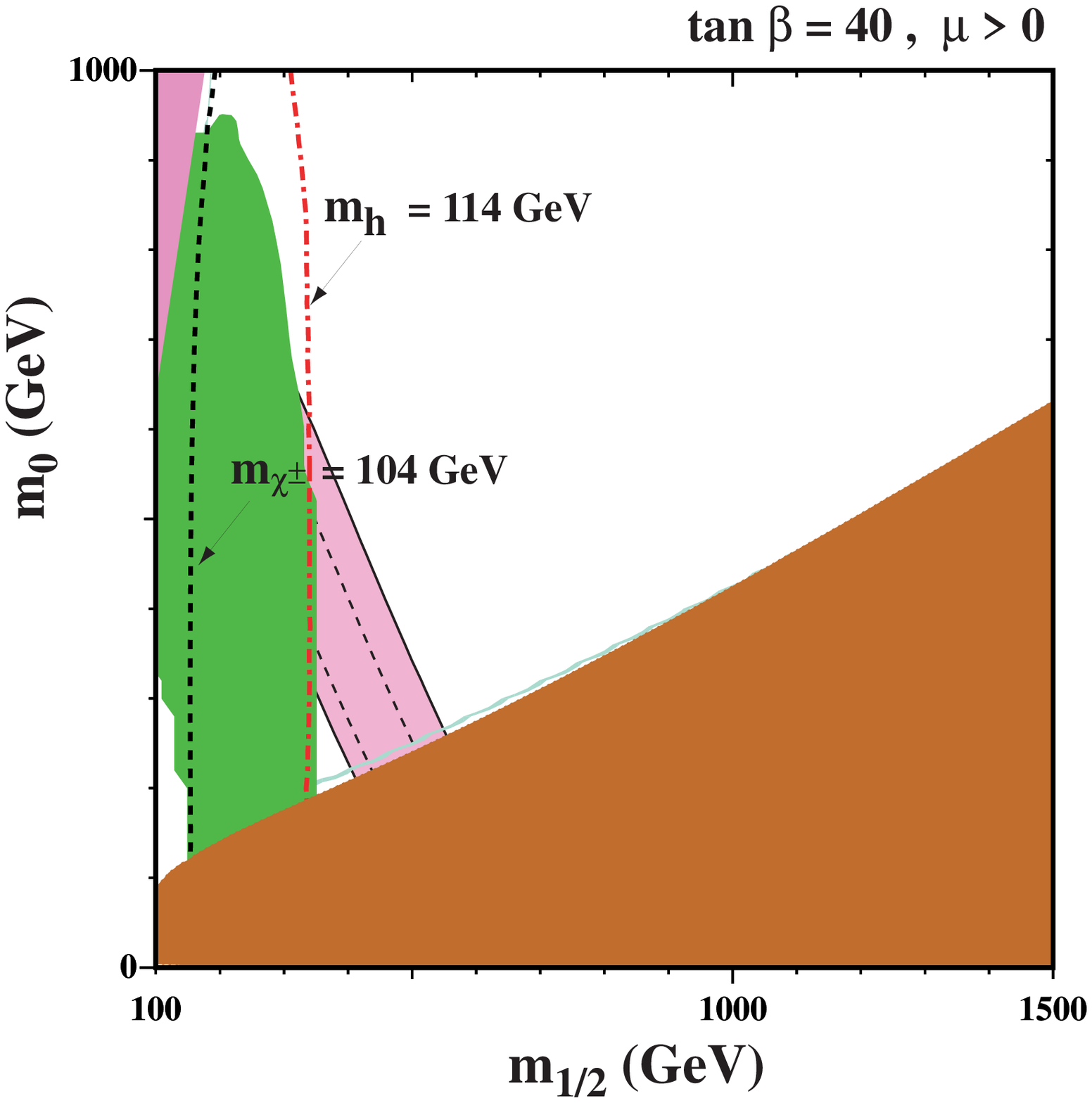}}\subfigure[]{\includegraphics[width=.33\textwidth]{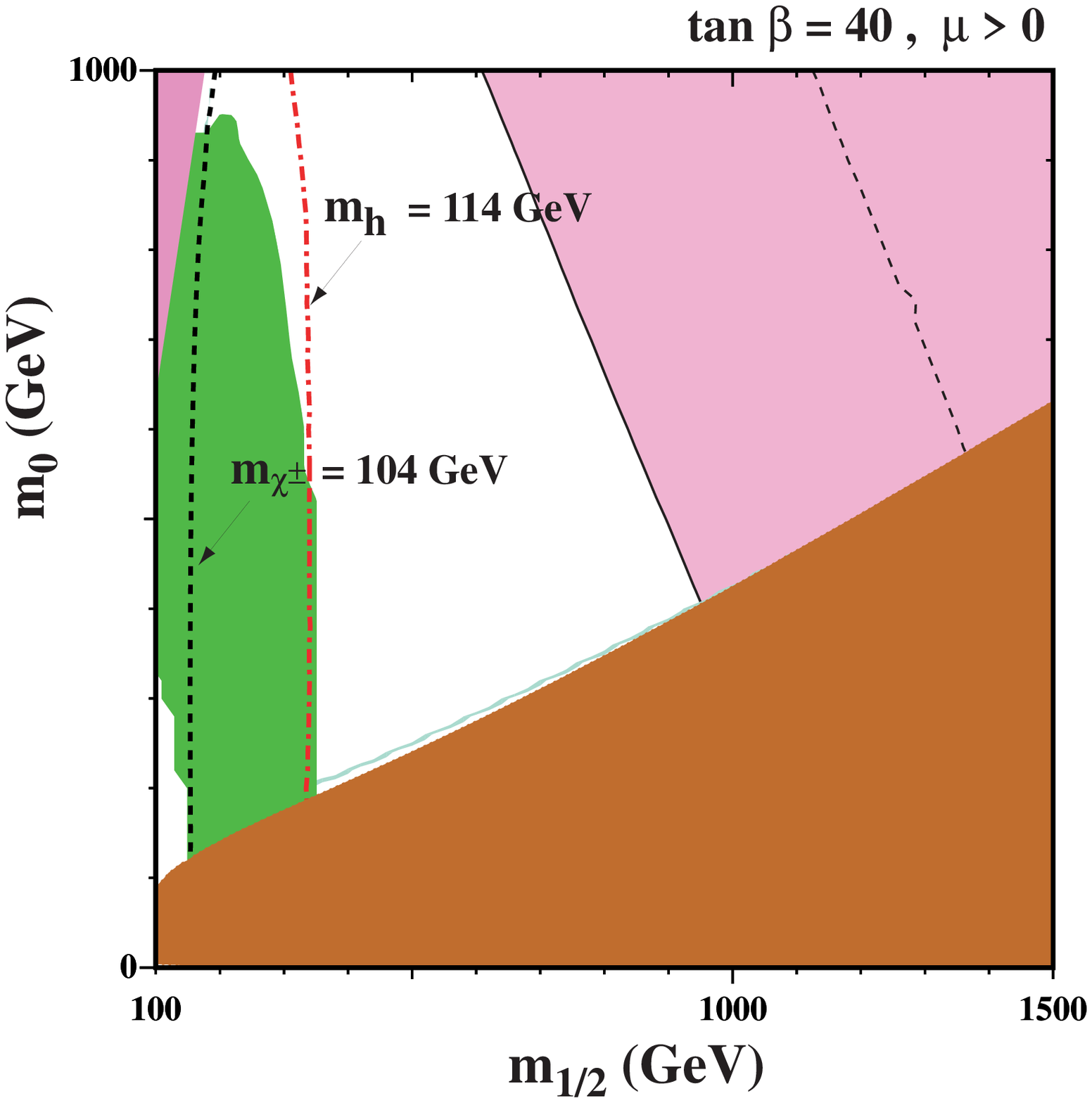}}
\caption{The CMSSM plots as above, but 
with $\tan \beta = 40$.
(a) As in Fig.~\ref{fg:dark} but for $\tan \beta = 40$ (b) The plot with
 $\Delta a_{\mu}=295(39)\times 10^{-11}$, which assumes that both
the theory and experimental errors decrease to $22 \times 10^{-11}$.
(c) The same errors as (b), but $\Delta = 0$.
 (Figures courtesy of K. Olive)
\label{fg:dark2}}
\end{center}
\end{figure}

A third example concerns the restriction of special, highly constrained models
of new physics such as the constrained MSSM (CMSSM)~\cite{olive}. 
The CMSSM has only four
free continuous parameters. One precise measurement such as the future
determination of $\Delta\amu$ effectively fixes one parameter as a function of
the others and thus reduces the number of free parameters by one.
In fact, the CMSSM is very sensitive not only to the $\amu$ but
also to the dark matter (assumed to consist of neutralinos) relic
density. As shown in Figs.~\ref{fg:dark} and \ref{fg:dark2}, 
both observables lead to
orthogonal constraints in CMSSM parameter space, and therefore imposing both
constraints leaves only two free parameters and thus allows for very stringent
tests of the CMSSM at the LHC.

\subsection{$\amu$ is sensitive to quantities that are difficult to measure at
  the LHC}

The LHC as a hadron collider is particularly sensitive to colored particles
whereas the $\amu$ measurement is particularly sensitive to weakly interacting
particles that couple to the muon. Therefore the sensitivities are
complementary. As an example, if the MSSM is realized it is possible that the
LHC finds some but not all superpartners of the gauge and Higgs bosons, the
charginos and neutralinos. 

Furthermore, for unraveling the mysteries of TeV-scale physics it is not
sufficient to determine which kind of new physics, i.e.\ extra dimensions,
supersymmetry or something else, is realized, but it is necessary to determine
model parameters as precisely as possible. In this respect the complementarity
between the LHC and $\amu$ becomes particularly important. 
A difficulty at the LHC is the very
indirect relation between LHC observables (cross sections, mass spectra,
edges, etc) and model parameters such as masses and couplings, let alone more
underlying parameters such as supersymmetry-breaking parameters or the
$\mu$-parameter in the MSSM. It has been shown that a promising strategy is to
determine the model parameters by performing a global fit of a model such as
the MSSM to all available LHC data. However, recent investigations have
revealed that in this way typically a multitude of almost degenerate local
minima of $\chi^2$ as a function of the model parameters results
\cite{PlehnRauch}. Independent observables such as $\amu$ will be highly
valuable to break such degeneracies and in this way to unambiguously determine
the model parameters.

In the following we discuss the complementarity of LHC and $\amu$ for the
well-studied case of the MSSM, where it has turned out that the LHC has a weak
sensitivity to two central parameters: The LHC has virtually no sensitivity at
all to the sign of the $\mu$-parameter and only a moderate sensitivity to
$\tan\beta$, the ratio of the two Higgs vacuum expectation values.

The MSSM contributions to $\Delta\amu$ on the other hand are highly sensitive
to both of these parameters,
\begin{equation}
\Delta\amu^{\rm MSSM} \approx 130\times 10^{-11}\ \tan\beta\ \mbox{sign}(\mu)
\ \left(\frac{100\ \rm GeV}{M_{\rm SUSY}}\right)^2,
\end{equation}
where $M_{\rm SUSY}$ denotes the average superpartner mass scale. Therefore,
a future improved $\amu$ measurement has the potential to establish a
definite positive or negative sign of the $\mu$-parameter in the MSSM, which
would be a crucial additional piece of information. 

In order to discuss the relative sensitivity of LHC and $\amu$ to $\tan\beta$,
we reconsider the situation discussed in Ref.\ \cite{LaPlZerwas}. In this 
reference
it has been assumed that the MSSM reference point SPS1a is realized, and the
potential of the LHC to determine MSSM parameters has been worked out. By
performing a global fit of the MSSM to all available LHC data, a large set of
MSSM parameters can determined to a precision of a few percent. Apart from
sign$(\mu)$, which has been assumed to be positive,  $\tan\beta$ could be
determined only poorly to $\tan\beta^{\rm LHC\ fit}=10.22\pm9.1$. 

\begin{figure}[h]
\begin{center}
\includegraphics[width=0.45\textwidth]{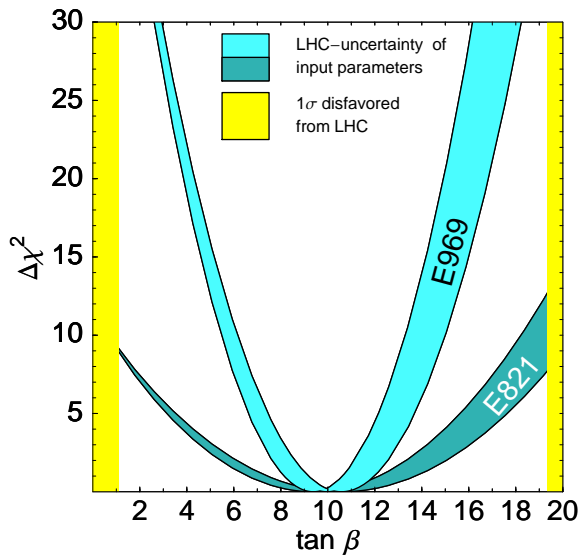}
\caption{A possible future ``blue band'' plot, where $\tan\beta$ is determined
from the measurement of $\amu$. The white region between the
yellow vertical bars indicates 
the $\pm 1\sigma$ region from the LHC-determination of
$\tan\beta^{\rm LHC\ fit}=10.22\pm9.1$. 
The darker blue band is with the present 
E821 restrictions.  The lighter blue band corresponds to 
$\Delta\amu^{(\rm future)}=295(39)\times 10^{-11}$. 
 It is assumed that the MSSM reference point
SPS1a is realized and that the MSSM parameters have been determined by a
global fit to the values given in Ref.\ \cite{LaPlZerwas}, Table~5,
and that the
measured value $\amu^{\rm exp}$ coincides with the actual value of the SPS1a
point.  The plot shows $\Delta\chi^2
=\left(\frac{\amu^{\rm MSSM}(\tan\beta)-\amu^{\rm
      exp}}{\{88;39\}\times 10^{-11}}\right)^2$ as a 
function of $\tan\beta$, where in $\amu^{\rm MSSM}(\tan\beta)$ all parameters
except $\tan\beta$ have been set to the values determined at the LHC. The
width of the blue curves result from the uncertainty of these parameters. 
The plot shows that the precision for
$\tan\beta$ that can be obtained using $\amu$ is limited by the precision of
the other input parameters but is still better than $20\%$ and thus much
better than the determination using LHC data alone.
\label{fig:blueband}
}
\end{center}
\end{figure}

In such a situation, an improved $\amu$ measurement will be the perfect
complement of the LHC. One can simply study the MSSM-prediction for
$\Delta\amu$ as a function of $\tan\beta$ (all other parameters are known from
the global fit to LHC data) and compare it to the measured value. One can
display the result in a ``blue band'' plot, similar to the case of the LEP
precision data, which can be compared to the standard model predictions as a
function of the standard-model Higgs boson mass. The resulting possible future
``blue band'' plot for $\tan\beta $ determined by the $\amu$ measurement is
shown in Fig.~\ref{fig:blueband}. As can be seen from the plot, the
improvement of the determination of $\tan\beta$ from the $\amu$ measurement is
excellent.

One should note that even if better ways to determine $\tan\beta$ at the LHC
alone might be found in the future,  an independent determination using $\amu$ 
will still be highly valuable. $\tan\beta$ is one of the central MSSM
parameters, and it appears in all sectors and in almost all
observables. Therefore, measuring $\tan\beta$ in two different ways,
e.g.\ using certain Higgs- or $b$-decays at the LHC and 
using $\amu$, would constitute a non-trivial and indispensable test of the
universality of $\tan\beta$ and thus of the structure of the MSSM. 

\subsection{$\amu$ is an inclusive measure of quantum effects}

At the LHC, it is not trivial to discover all new particles that are in
principle kinematically accessible. Some, in particular unexpected particles,
might be difficult to detect due to background problems, not-optimized search
strategies, or the triggers. The quantity $\amu$ on the other hand, as a pure
quantum effect, is sensitive to all particles that couple to the muon (and,
from the two-loop level on, even to particles that don't couple to the
muon). Therefore, a precise measurement of $\amu$ constitutes an inclusive
test of all kinds of new physics, expected or unexpected. 

If a large deviation from the standard model prediction is found, this can
help to establish the existence of new particles that have not been seen at
the LHC. The projected precision of the $\amu$ measurement will 
even permit the derivation of
mass bounds on such new particles. Feeding this information back to the
LHC will help optimizing searches and might thus make a direct detection of
the new particles possible.

Likewise, if a small $\Delta\amu$ is found, this will help exclude the
existence of certain particles in a particular mass range. In this way,
regions of parameter space that are difficult to test at the LHC might be
covered. 10 years ago, LEP-experiments could not exclude the existence of
particularly light charginos, since they could have escaped detection. This
hole in parameter space was then closed by considering $\amu$
\cite{WagnerLEP}. Such light charginos would have given a large contribution
to $\amu$ which was not observed. 

\subsection{$\amu$ is a clean observable}

The LHC is an extremely complex machine, and it will be a huge task to
understand the LHC detectors sufficiently to make reliable measurements
possible. At the LHC many sources of systematic errors have to be brought
under control, and the overwhelming background makes it difficult to extract
meaningful signals. The $\amu$ measurement suffers from none of these
problems. 

Therefore, the errors associated with LHC and the $\amu$ measurement are
totally complementary, and the $\amu$ measurement will constitute a
non-trivial cross-check of the LHC. The importance that LHC performs
successfully cannot be overestimated. This implies that independent
measurements that can cross-check and guide the LHC with clean data are
equally important.

\subsection{$\amu$ is a simple and beautiful observable}

Finally, it is worth mentioning that the role the observable $\amu$ has played
in the past and should play in the future goes far beyond just being a useful
tool. $\amu$, and anomalous magnetic moments in general, are some of the
simplest and most beautiful observables in fundamental physics. They have
found entrance in all quantum field theory textbooks and have inspired
generations of quantum field theory students and researchers.

Anomalous magnetic moments are the simplest observables for which quantum
effects in quantum field theory are important. The first measurement of the
anomalous magnetic moment of the electron sparked the first successful
loop calculation in QED by Schwinger, in the course of which the basic ideas
of renormalization theory were developed. In the meantime, many more
milestones in the understanding of quantum field theory are related to, 
and were inspired by research on anomalous magnetic moments.

Furthermore, $a_\mu$ has great appeal to the general public.  All of the E821
results were covered by the {\it New York Times}
and the rest of the popular
press, as well as other journals such as  {\it Science News},
{\it The New Scientist},  {\it Physics Today},
  {\it Science}, and {\it Nature}.
Measuring a quantity with such high precision that one can resolve effects 
from almost all elementary particles, ranging from the photon, electron,
muon, over hadrons, to $W$ and $Z$ bosons, is striking and catches the
imagination.  The projected precision of the measurement will 
permit the resolution of
effects from new particles such as supersymmetric particles.  This
opportunity should not be missed.

%% file: improvements.tex
\section{Improvements to the $(g-2)$ Experiment}

The final error of 0.54~ppm obtained in E821 was statistics limited,
with a 0.46~ppm statistical error and a 0.28~systematic error.  The
errors from each running period are given in
Table~\ref{tb:error-all}.
Any upgraded experiment must further improve the systematic
 errors  and significantly increase
 the volume of data collected.  The principal focus
of this document is to present the {\it physics case} 
for an improved experiment, rather than  to present technical 
details of the upgraded experiment. We give no specifics for experimental
 improvements in this white paper, but rather
briefly describe the experiment and  the possible future goals.

\begin{table}[h]
\begin{center} \caption{Systematic and statistical errors
in ppm for each of the e821 running periods. } \label{tb:error-all}
\begin{tabular}{|l|c|c|c|c|c|}
\hline  &  1998 &  1999 &  2000 &  2001 &
E969 Goal\\ \hline Magnetic Field Systematic ($\omega_p$) & 0.5 &  0.4
&  0.24 &  0.17 &  0.1
\\
\hline Anomalous Precession Systematic ($\omega_a$)\ \ & 0.8 & 0.3 & 0.31
&  0.21 &
0.1 \\
\hline
Statistical Uncertainty & 4.9 &   1.3 &   0.62 &  0.66 &  0.14 \\
\hline
Total Uncertainty    &     5.0 &   1.3 &   0.73 &  0.72 &  0.20 \\
\hline
\end{tabular}
\end{center}
\end{table}

A proposal to Brookhaven,  E969, which received enthusiastic
scientific approval, plans to reduce the combined error to
0.2~ppm, a factor of 2.5 improvement beyond E821.  When combined
with expected improvements in the strong-interaction contribution to
the standard-model value, the improved sensitivity could increase
the significance of any difference between theory and experiment to
above the $6~\sigma$ level, assuming the central values remain the same.
  Some members of the community have
encouraged a close look to see if a factor of 5
improvement  is possible, down
to a precision of $\pm 0.1$~ppm.

\subsection{How is $(g-2)_\mu$ measured?}

The muon anomalous moment is determined from the difference
frequency \wa\ between the spin precession frequency,
$\omega_S$, and the cyclotron frequency, $\omega_C$,
of an ensemble of polarized muons that circulate in a storage ring
having a highly uniform magnetic field. Apart from very small
corrections, \wa\ is proportional to \amu. Vertical containment 
in the storage ring is
achieved with an electric quadrupole field.
In the presence
of  magnetic and electric fields having $\vec {B} \cdot \vec {\beta}
= \vec {E} \cdot \vec {\beta} = 0$, \  $\omega_a$ is described by
\be
\vec \omega_a = \vec \omega_S - \vec \omega_C = 
 - \ \frac{q }{ m} 
\left[ a_{\mu} \vec B -
\left( a_{\mu}- \frac{1 }{ \gamma^2 - 1} \right)
\frac{ {\vec \beta \times \vec E }}{ c }
\right]\, ,
\label{eq:spin-precess}
\ee
where $q = \pm e$ is the muon charge and $\vec{\beta}$ is the 
muon velocity in units of $c$. The
term in parentheses multiplying $\vec{\beta} \times \vec{E}$
vanishes at the ``magic'' value of $\gamma = 29.3$, 
and the electrostatic focusing does not
affect the spin motion
(except for a small correction necessary to account for
the finite momentum range $\Delta P/P \approx \pm0.14\%$ around the
magic momentum).

\begin{figure}[h]
\includegraphics[width=\textwidth]{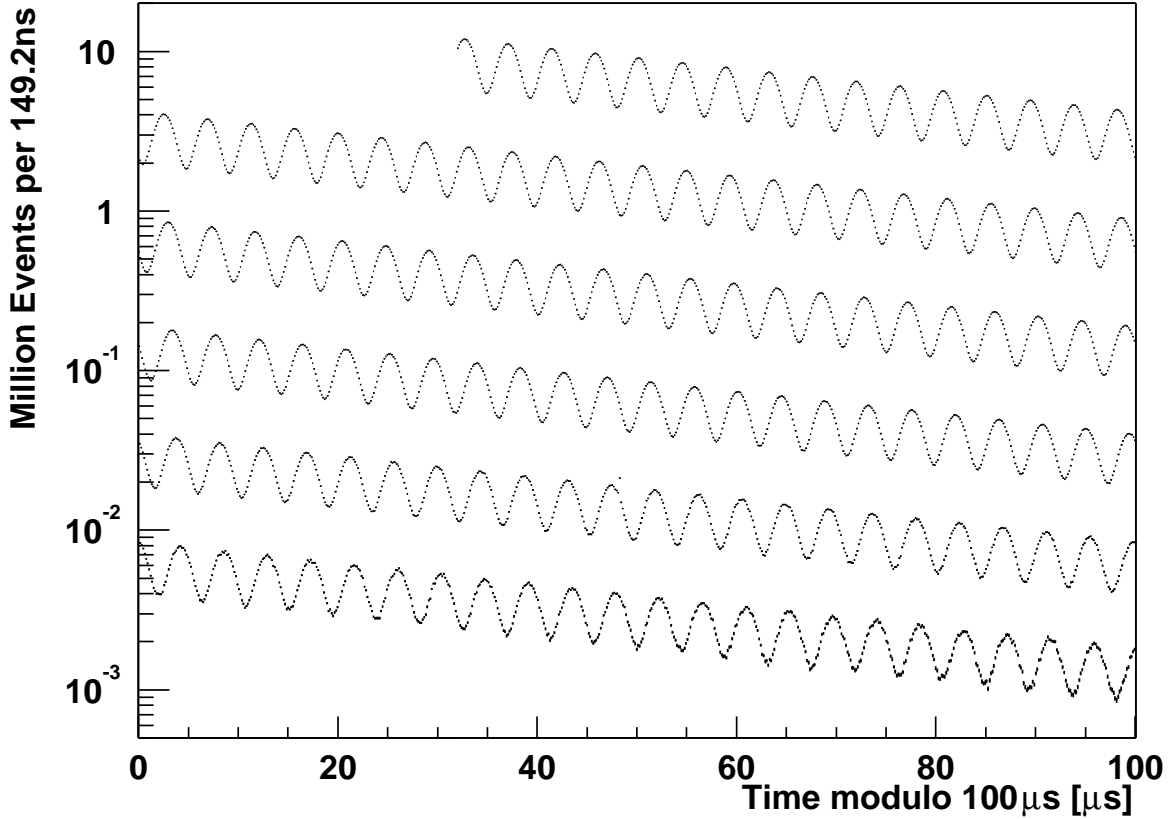}
\caption{The time spectrum of $3.6 \times 10^9$ 
electrons with energy greater than
1.8 GeV from the 2001 data set. 
The diagonal ``wiggles'' are displayed  modulo 100~$\mu$s.
Data are from Ref.~\cite{bennett3}.
\label{fg:wiggles}}
\end{figure}

The magic momentum of 3.094~GeV/c sets the scale of the experiment, and the BNL
storage ring~\cite{ring} is 7.1~m in radius and has a 1.45~T
magnetic field. At 3.094~GeV/$c$ the time-dilated muon lifetime is
64.4~$\mu$s, and the decay electrons
have a maximum lab-frame energy of approximately 3.1~GeV.
A short ($\sim 25$~ns)
bunch of muons is injected into the storage ring, and the arrival time
and energy of the decay electrons is measured.
The time spectrum of muon decay electrons above a single energy threshold
 produces the time distribution 
\begin{equation}
N(t,E_{th}) = 
N_{0}(E_{th})e^\frac{-t}{\gamma\tau}
[1+A(E_{th})\cos(\omega_{a}t+\phi(E_{th}))] \, ,
\label{eq:fivep} 
\end{equation}
as shown in Fig.~\ref{fg:wiggles}.
The value of $\omega_a$ is obtained from a 
least-squares fit to these data.  

In the experiment the muon frequency $\omega_a$  is determined
to high precision, and the average magnetic field is measured to equal or
better precision.  The field is determined from a suite of NMR
measurements~\cite{bennett3,nmr}: 
to reference the field against an absolute standard; to monitor
the field continuously; and to map the field in the storage ring
aperture~\cite{bennett3}. 

An upgraded experiment at any level requires a significant increase
in the muon beam intensity, as well as improvements in the detectors and 
front-end electronics.  A credible case 
was made in the E969 proposal that the factor of 
2.5 improvement could be realized.  We believe that with further 
research and development, and adequate running time on the 
accelerator, a significant increase in precision beyond the 
factor of 2.5 could be achieved.  We continue to study potential
improvements to the beamline, and to the electron detectors and electronics.

%% file: Summary.tex
\section{Summary}
In this White Paper, we concentrate on the physics case for the new
\gm\ experiment, E969.  The standard model theory situation
now---Spring, 2007---gives a precision commensurate with experiment
at roughly 0.5~ppm.  Improvements are expected from ongoing
work, both in the experiment-driven lowest-order
hadronic vacuum polarization, and in the hadronic light-by-light contributions.
 The new $(g-2)$ experiment will reduce the experimental
error  by a factor of
2.5 (or more) so that a comparison with theory will have a combined
sensitivity to $\Delta a_\mu$ better than $47 \times
10^{-11}$.  The current discrepancy, at $3.4~\sigma$ would rise
above $6~\sigma$ if the magnitude of the difference remains
unchanged.
New physics signals are expected to emerge in the LHC era, and the
improved \amu\ measurement can make a significant impact in 
the role of helping to sort out the nature of the discoveries
made at the LHC.  The precision
physics from low-energy observables is complementary to the
discovery potential of the collider.  Many authors have understood
the importance of the independent constraint placed on 
new physics by $(g-2)_\mu$, and there are over
1300 citations to the E821 papers.  

We conclude with a list of items related to ``Why now?"

\begin{itemize}

 \item The experimental precision can be improved by at least a factor 
of 2.5 or more but it must
be started ``now'' to be ready when the results from the LHC will
demand additional constraints. Several years of R$\&$D and
construction are required before running and analysis can begin.  We
estimate roughly $5 - 6$ years from project start to achieve the
goal.

  \item The standard model theory uncertainty is slightly smaller
  than experiment and it should be halved again over the next few years.
  Over the past twenty years E821 stimulated an
enormous amount of theoretical work, which necessitated the breaking of
new ground in higher-order QED and electroweak contributions, as well as
the significant work on the hadronic contribution.
These improvements have been driven by the fact that real measurements
  of \amu\ were also being made.  The momentum should be sustained and
  new efforts, especially related to the difficult hadronic
  light-by-light contribution, must be encouraged.

  \item We are already at a compelling moment.  The present 
  $e^+e^-$-based standard model 
   theory is $3.4$ standard deviations from the experiment, providing
  a strong hint of new physics. If the current discrepancy
  persists, with halved experimental and theoretical
  uncertainties, the significance will rise above $6~\sigma$.

  \item For specific models, such as SUSY, \amu\ is particularly effective at
  constraining $\tan\beta$---the ratio of Higgs vacuum expectation 
values---for
  a given superparticle mass and it gives the sign of the $\mu$ parameter,
  something that cannot be obtained at the LHC.  
This information is complementary to the
  anticipated LHC new-particle spectra and it will
 be crucial in the effort to pin down the
  parameters of the theory behind them.

  \item Independently of SUSY---we do not suggest or depend on this or
  any other specific model as being correct---measuring \amu\ to
  very high precision will register an important constraint for any new
  physics theory to respect.  Some models will predict a large \gm\
  effect, while others will not.  It is information that can likely
  help diagnose new physics.

  \item On the practical side, the project is based on a proven track record by
  an existing team of experts and new enthusiastic collaborators.
  It is efficient to mobilize the Collaboration now while the storage
ring and beamline
  facilities can be dependably re-commissioned, and while the diverse
  expertise exists.
\end{itemize}

\vskip 0.3in
\noindent {\it Acknowledgments: We thank Michel Davier
and Simon Eidelman for their helpful comments and discussion on the hadronic
contribution.  We thank Keith Olive for providing the CMSSM figures.
}